\documentclass[aps,prb,twocolumn,superscriptaddress]{revtex4-1}
   
\usepackage{amsmath}
\usepackage{amssymb}
\usepackage{graphicx}
\usepackage{physics}
\usepackage{color}
\usepackage{bm}

\begin{document}
\newcommand{\jiajun}[1]{\textcolor{blue}{#1}}
\newcommand{\denis}[1]{\textcolor{green}{#1}}
\newcommand{\I}{\mathrm{i}}

\title{$\eta$--paired superconducting hidden phase in photodoped Mott insulators}
\author{Jiajun Li}
\affiliation{Department of Physics, University of Erlangen-Nuremberg, 91058 Erlangen, Germany}
\author{Denis Golez}
\affiliation{Center for Computational Quantum Physics, Flatiron Institute, 162 Fifth Avenue, New York, NY 10010, USA}
\author{Philipp Werner}
\affiliation{Department of Physics, University of Fribourg, 1700 Fribourg, Switzerland}
\author{Martin Eckstein}
\affiliation{Department of Physics, University of Erlangen-Nuremberg, 91058 Erlangen, Germany}

\date{\today}

\begin{abstract}
We show that a metastable $\eta$--pairing superconducting phase can be induced by photodoping doublons and holes into a strongly repulsive fermionic Hubbard model. The doublon-hole condensate originates from an intrinsic doublon-hole exchange interaction and does not rely on the symmetry of the half-filled Hubbard model. It extends over a wide range of doublon densities and effective temperatures. Different non-equilibrium protocols to realize this state are proposed and numerically tested. We also study the optical conductivity in the superconducting phase, which exhibits ideal metallic behavior, i.e., a delta function at zero-frequency in the conductivity, in conjunction with a negative conductivity at large frequencies. These characteristic optical properties can provide a fingerprint of the $\eta$-pairing phase in pump-probe experiments.
\end{abstract}

\maketitle

\section{Introduction} 
Non-equilibrium phenomena hold the promise of creating new phases of matter  
and selectively enhancing different orders \cite{basov2017,ichikawa2011,stojchevska2014,mor2017}. 
One of the most tantalizing findings in the field is the possible light-induced superconductivity in strongly correlated materials \cite{fausti2011,mitrano2016}. Various theoretical works have attempted to explain its microscopic origin in different situations \cite{raines2015,denny2015,patel2016,okamoto2016,babadi2017,murakami2017,kennes2017,mazza2017,matthies2018}. In addition to ideas based on Floquet engineering of electron-phonon and electron-electron interactions, one may contemplate the interesting possibility that $\eta$--\emph{pairing} plays a role in these scenarios. 

The $\eta$--paired states are a family of excited states of the fermionic Hubbard model on a bipartite lattice, which exhibit an unconventional staggered superconducting
order parameter \cite{yang1989}. A plethora of non-equilibrium protocols have been considered to selectively %pick up 
induce
the $\eta$--paired states. In the ideal Hubbard model, the $\eta$--order is related to spin and charge orders within the $SO(4)$ symmetry, and can be induced out of the charge-density-wave or $s$--wave superconducting ground state in the attractive Hubbard model \cite{yang1990, demler2004, sentef2017, kitamura2016}. 
Recently, carefully designed non-equilibrium protocols have been shown to populate the $\eta$--paired states and selectively suppress competing antiferromagnetic correlation in the \emph{repulsive} Hubbard model \cite{kaneko2019,tindall2019, bernier2013}. This finding opens up the tantalizing possibility of inducing superconductivity in a Mott insulator, and is related to recent experiments on light-induced superconductivity \cite{tindall2020}. However, these works did not suffice to conclude a symmetry-breaking phases of $\eta$--pairing, which should be indicated by a \emph{divergent} pairing susceptibility (so an infinitesimal perturbation can induce a growing SC phase domain in an extended system), and, furthermore, require stringent conditions on the external driving as well as the $SO(4)$ symmetry of the half-filled Hubbard model, which is often broken in real materials, resulting in, e.g., the decay of the pumped $\eta$--pairing \cite{kaneko2019}. Therefore, an intriguing question arises whether nonequilibrium protocols can induce a metastable $\eta$--paired \emph{hidden phase} which is robust against symmetry-breaking perturbations.

In recent years, photodoping has emerged as one of the most promising pathways to induce nonthermal phases in strongly correlated materials \cite{ichikawa2011,stojchevska2014}. 
Here, we use the term photodoping to refer to any non-equilibrium protocol that creates charge carriers
%Here, we refer by photodoping to any non-equilibrium protocol creating excess charge excitations 
in an insulating system (in particular a Mott insulator). Because photocarriers can have a long lifetime  \cite{iwai2003, rosch2008, okamoto2010, sensarma2010, eckstein2011, lenarcic2013, mitrano2014}, fast intraband thermalization processes can eliminate detailed memory of the nonequilibrium protocols and lead to a \emph{universal photodoped state} characterized by only few parameters, such as the doublon density $d$ and an effective temperature $T_{\rm eff}$. The partial thermalization can be particularly efficient with the recently proposed evaporative cooling mechanism \cite{werner2019}. So far, weak photodoping of doublon-hole pairs into the half-filled Mott state has been observed to slightly enhance the local pairing susceptibility \cite{werner2018,peronaci2019}. However, except for the extreme limit $d=0.5$, in which all sites contain either doublons or holons \cite{rosch2008}, it remains unclear how much the $\eta$--pairing can be enhanced, and whether an $\eta$--paired hidden phase can be stabilized upon photodoping.

In this Article, we show that photodoping can indeed induce a robust hidden phase with $\eta$--pairing in the Mott insulator for a wide 
range of parameters $d$ and $T_{\rm eff}$, by considering exemplarily a Hubbard lattice coupled to external fermion reservoirs. A steady-state 
dynamical mean-field theory is used to solve the problem and scan a \emph{non-equilibrium phase diagram}. The hidden phase 
originates from a doublon-hole pairing mechanism that is intrinsic to the local electron-electron interaction. The instability only relies on the presence of cold 
photocarriers and requires no special photodoping protocols, or the $SO(4)$-symmetry protection. We further demonstrate that this state behaves like a superconductor, i.e., it shows zero resistivity and the Meissner effect.

The Article will be orgainzed as follows. Sect.~II discusses the general concept of a quasi-stationary photodoped state in a single-band Mott insulator. Sect.~III introduces the steady-state formulation and the numerical method (dynamical mean-field theory). Sect.~IV shows the numerical results and the analytical understanding of $\eta$--paired superconductivity in phodotoped Mott insulators. Sect.~V includes conclusion and outlook.

\section{Photodoping of a Mott insulator} 
We consider the repulsive Hubbard Hamiltonian on a bipartite lattice at half-filling ($\langle n\rangle=1$),
\begin{align}
H=-t_0\sum_{\langle ij\rangle\sigma} d^\dag_{i\sigma} d_{j\sigma}+U\sum_{i}n_{i\uparrow} n_{i\downarrow},
\label{ham}
\end{align}
with hopping $t_0$ and interaction $U$; $t_0=1$ sets the energy scale in the following. 

The model \eqref{ham} features a Mott insulating ground state with strong antiferromagnetic correlations at half-filling. The ground state at half-filling is generically non-superconducting. To induce non-zero $\eta$-pairing, one has to drive the system out of equilibrium. In strongly correlated solids, one way
 is to excite the system with electromagnetic waves, creating charge carriers across the insulating gap \cite{eckstein2011,werner2012}. 
 This leads to a photo-doped state, as outlined in the introduction. 
 Here we propose that  any general protocol which creates charge excitations in the system %without referring to the specific form of electromagnetic coupling, should show 
 should result in 
 similar physics %as the photo-doped 
 sufficiently long after the excitation itself. 
 This is justified by the hierarchy of time scales: in general, the charge recombination process is 
 relatively slow, because the dissipation of a large potential energy $U$ to low-energy degrees of freedom is inefficient 
\cite{sensarma2010,sensarma2011}, while the intra-band thermalization (doublon-doublon and holon-holon scattering) can be much faster. Thus, a partially thermalized photodoped state, 
 with metastable doublons and holons as quasiparticles, can be observed in a reasonable time window.

\section{Model and method }
This fact motivates us to consider a steady-state bath-coupling mechanism as a representative of general photodoping protocols. The details of steady-state theory for photodoping can be found in a recent work \cite{li2020ness}. We will first discuss this mechanism and compare the results against other protocols. To be specific, we consider a coupling to auxiliary external fermion baths at each lattice site through %quadratic coupling, 
\begin{align}
H_{\rm bath}&=\sum_{i\alpha\sigma}\varepsilon_\alpha c_{i\alpha\sigma}^\dag c_{i\alpha\sigma}+g\sum_{i\sigma\alpha}d^\dag_{i\sigma}c_{i\alpha\sigma},
\end{align}
where $g$ is the coupling constant between lattice electrons $d_{i\sigma}$ and bath electrons $c_{i\alpha\sigma}$ and $\alpha$ labels different baths and bath levels of energy $\varepsilon_\alpha$. We take into account two separate fermion baths of half-bandwidth $W=2$ with hybridization density of states $D(\omega)_{\pm}=\sum_{\alpha}g^2\delta(\omega-\epsilon_\alpha)=\Gamma\sqrt{W^2-(\omega\pm U/2)^2}$, where $\Gamma=g^2/W^2$. The bath $D_\pm$ is shifted by $\pm U/2$, and the corresponding chemical potential is shifted by $\pm\mu_b$. Such a bath coupling results in the injection of electrons into the upper Hubbard band and the absorption of electrons out of the lower Hubbard band. Due to the long lifetime of the excess doublons and holes, large doublon occupancies $d=n_\uparrow n_\downarrow$ can be reached with relatively small bath couplings $\Gamma$, minimizing the side effects of the bath coupling, so that the behavior of the non-equilibrium steady-state reflects the general properties of photodoped systems.

The model is considered on the infinitely coordinated Bethe lattice with non-interacting bandwidth $4t_0$, where it can be exactly solved using non-equilibrium Dynamical Mean-Field Theory (DMFT) \cite{georges1996,aoki2014}, both for the real time dynamics under time-dependent driving protocols, and for non-equilibrium steady states. The Bethe lattice is bipartite and can be used to study symmetry-breaking phases with either uniform or staggered order parameters. In the staggered case, the lattice model is mapped to two self-consistent Anderson impurity models defined by $S_{A/B}[\psi,\bar{\psi}]=S^{\rm loc}_{A/B}[\psi,\bar{\psi}]-\int dt dt' \bar{\psi}(t)\Delta_{A/B}(t,t') \psi(t')$, which are solved using the non-crossing approximation \cite{eckstein2010}. The resulting DMFT equations in the Nambu Keldysh formalism are similar to those of staggered antiferromagnetism and $s$--wave superconductivity, with the hybridization function $\Delta_{A/B}(t,t')=t_0^2\sigma_z G_{B/A}(t,t')\sigma_z+D(t,t')$. More details are given in the appendix.

\begin{figure}
\includegraphics[scale=0.7]{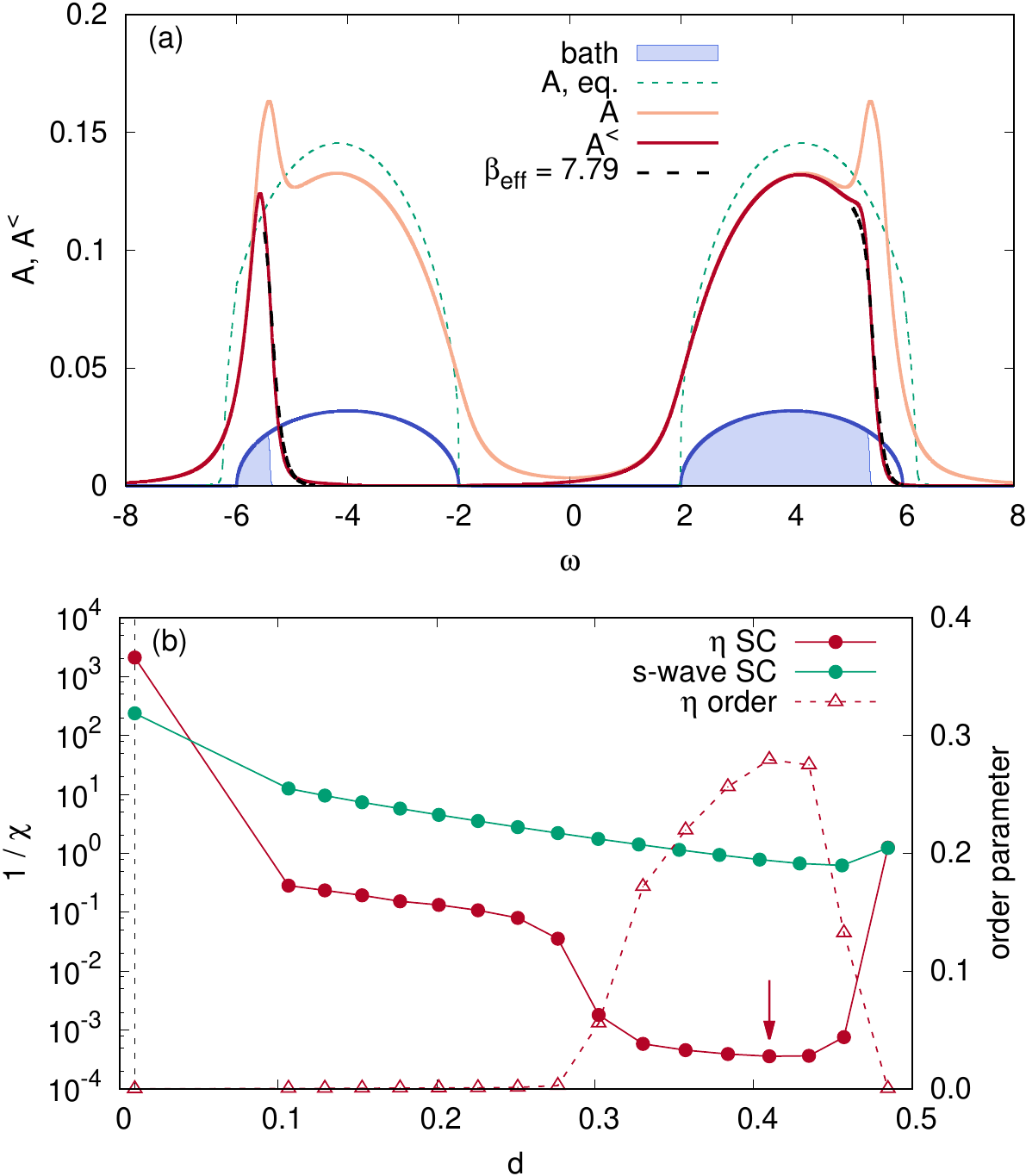}
\caption{(a) Spectral function $A(\omega)$ and occupied density of states $A^<(\omega)$ at $\mu_b$=5.4, which corresponds to the data point labeled by the arrow in (b). The dashed green curve indicates the equilibrium ($\mu_b=0.0$, $\Gamma=0.05$) spectral function for $\beta_{\rm eq}=100$. The blue curve shows the density of states of the baths, while their fillings at $\mu_b=5.4$ are shown as shaded areas. Dashed black lines indicate $A(\omega) f_{\rm FD}(\omega)$, with a Fermi distribution $f_{\rm FD}$ of inverse temperature $\beta_{\rm eff}=7.7901$ and chemical potential $\mu=\mu_b$. (b) Susceptibility of both $\eta$ and $s$--wave SC pairing as a function of double occupancy. $\Gamma=0.05$ and $\beta_b=100$. The equilibrium ($d\sim0.01$) is indicated by the vertical dashed line.}
\label{spec}
\end{figure}

The preparation of photodoped states via bath doping is illustrated in Fig.~\ref{spec}(a), for $U=8$ and $\Gamma=0.05$. The plot exemplarily shows the spectral function $A(\omega)$ and the occupied Density of States $A^<(\omega)=\operatorname{Im}G^<(\omega)/2\pi$ in the non-equilibrium steady state for one set of bath parameters. The curves can be related reasonably well by assuming a Fermi distribution function $f(\omega)=A^<(\omega)/A(\omega)$ with $\mu_b=\pm5.4$
at a given temperature (dashed line), thus verifying the universal nature of the bath-doped state which has been mentioned above. In the following, different doublon-hole densities and temperatures $T_{\rm eff}$ are fixed implicitly by varying $\mu_b$ and the bath temperature $T_b$. 

\section{superconductivity of the photocarriers} 
\begin{figure}
\includegraphics[scale=0.7]{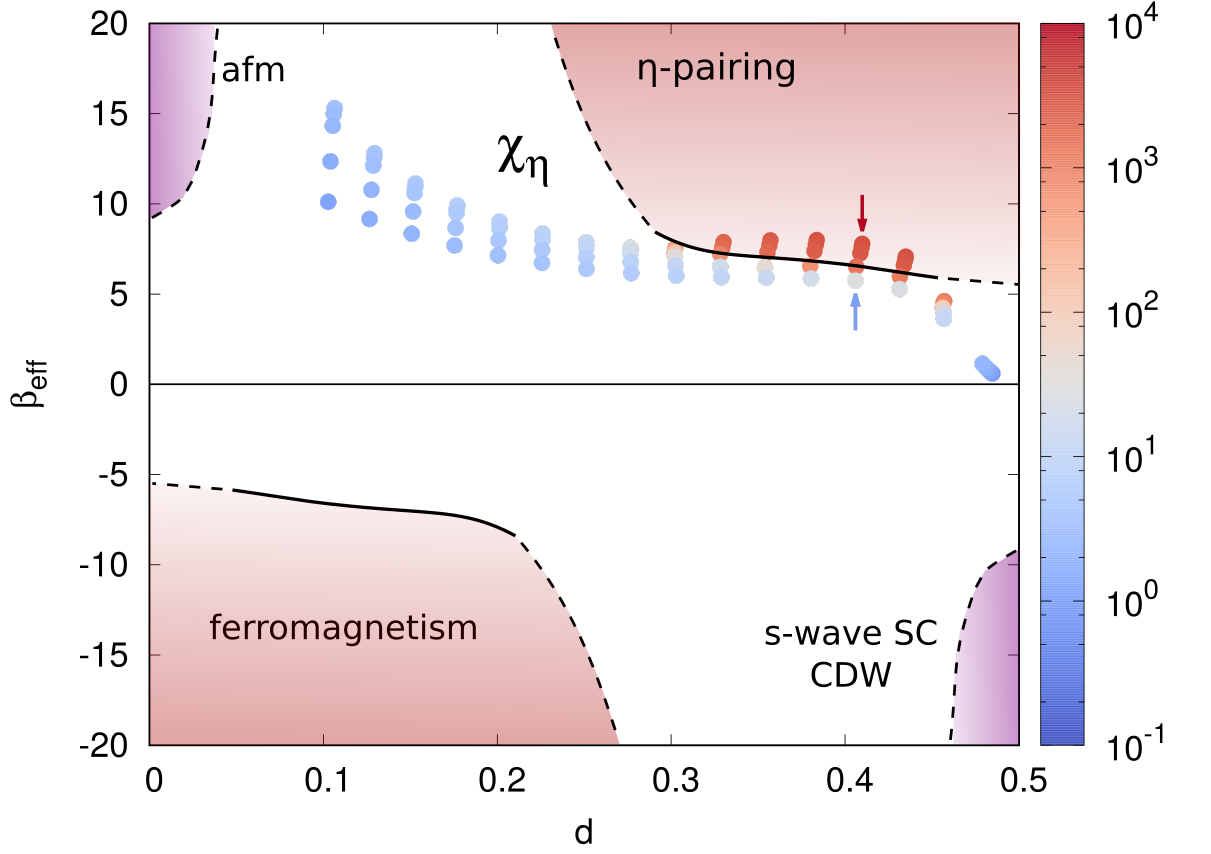}
\caption{Non-equilibrium phase diagram of the repulsive Hubbard model at $U=8$ under photodoping. The data points show the susceptibility $\chi_\eta$ along scans through the phase diagram, obtained by varying the inverse temperature of the auxiliary bath at $\Gamma=0.05$ and different $\mu_b$ from $\beta_b=100.0$, $50.0$, $33.3$, $20.0$, to $17.2$. The phase boundary is only schematic ($\chi_\eta\sim10^3$) and a guide to the eye. The negative temperature region is obtained from the positive one by reflection. The region close to equilibrium does not extend to $d=0$ but is limited by the double occupancy of the equilibrium state.}
\label{pd}
\end{figure}

To study the pairing susceptibility in the photodoped states, a local test field $\frac{1}{2}h_x (d_\downarrow d_\uparrow + {\rm h.c.})$ with $h_x=0.0001$ is applied to measure the pairing susceptibility $\chi=-\operatorname{Re}\langle d_\downarrow d_\uparrow\rangle/h_x$. Both uniform ($s$--wave) and staggered ($\eta$--) pairing susceptibilities are measured in the resulting photodoped states %The $\eta$ susceptibility in the resulting photodoped states 
for a scan with varying $\mu_b$, as plotted in Fig.~\ref{spec}(b). In the $\eta$--pairing case, the local test field is opposite for the two sublattices $A$ and $B$. The pairing susceptibility is generally enhanced for both $s$--wave and $\eta$--pairing, with the latter much more favored. A prominent observation is the emergence of a non-zero $\eta$--pairing order under strong photodoping $d\gtrsim0.3$. 
Note that with $h_x=0.0001$, there is a numerical limit of the order $1/h_x=10^4$ to the susceptibility, but an order parameter $\operatorname{Re}\langle d_\downarrow d_\uparrow\rangle\gtrsim0.2$ clearly indicates the symmetry breaking.
By also varying the inverse bath temperature $\beta_b$ and thus implicitly $\beta_{\rm eff}=1/T_{\rm eff}$, we obtain different scans which can then be combined into a phase diagram (Fig.~\ref{pd}), showing $\chi_\eta$ as a function of $d$ and $\beta_{\rm eff}$. The phase boundary between the normal and $\eta$--pairing phase around $d\gtrsim 0.3$ and $\beta_{\rm eff}\gtrsim 6.0$ can be roughly identified, except for very large doping $d\sim0.5$ or low temperature due to the difficulty of precisely controlling $\beta_{\rm eff}$ in these regimes. 

Close to the equilibrium half-filled state $d\sim0$ ($d\approx0.01$ for the shown parameters), we have also sketched the antiferromagnetic phase, which is known to persist for weak photodoping but is quickly destroyed due to the doublon (hole) hopping processes \cite{golez2014,balzer2015,grusdt2018}. (DMFT gives a stability range of $d\lesssim 0.05$ for the antiferromagnetic phase under photodoping in the same model \cite{werner2012}.) Apparently, the $\eta$--pairing phase persists under photodoping over a much larger doping range $d$ as compared to antiferromagnetism. 

\subsection{The universality of photodoped $\eta$--paired phases}
To explain the phenomenology, we first note that the $\eta$--pairing order parameter can be expanded into three pseudospin components 
spanning the charge-sector $SU(2)$ symmetry: $\eta_i^+=\eta_i^x+i\eta_i^y=\theta_i d^\dag_{i\uparrow} d^\dag_{i\downarrow}$, $\eta^-=(\eta^+)^\dagger$,  and $\eta_i^z=\frac{1}{2}(n_i-1)$, where $\theta_i=\pm1$ on the two sublattices. The $\eta$--pairing phase can then be explained by a superexchange mechanism between the $\eta$--pseudospins. In fact, for $U\gg t_0$, one can project out doublon-hole creation and recombination processes using a Schrieffer-Wolff transformation \cite{rosch2008,bukov2016}, and obtain a two-liquid model where a doublon-hole liquid with exchange interaction $-\sum_{\langle ij\rangle}J_{\rm ex}\boldsymbol{\eta}_i\cdot \boldsymbol{\eta}_j$ couples (through doublon/holon hopping) to a singlon liquid with AFM exchange interaction $\sum_{\langle ij\rangle}J_{\rm ex}\boldsymbol{S}_i\cdot \boldsymbol{S}_j$. Specifically, the effective Hamiltonian reads,
\begin{widetext}
\begin{align}
H^{\rm eff}=-\sum_{\langle ij\rangle}J_{\rm ex}\boldsymbol{\eta}_i\cdot \boldsymbol{\eta}_j+\sum_{\langle ij\rangle}J_{\rm ex}\boldsymbol{S}_i\cdot \boldsymbol{S}_j-t_0\sum_{\langle ij\rangle\sigma}[\overline{\mathcal{P}}_id^\dag_{i\sigma}d_{j\sigma}\overline{\mathcal{P}}_j+\mathcal{P}_i d^\dag_{i\sigma}d_{j\sigma}\mathcal{P}_j],
\label{heff_f0}
\end{align}
\end{widetext}
which includes both exchange interactions and a hopping term that ``exchanges" the position of a pair of neighboring doublon/holon and singlon. The operator $\mathcal{P}_i$ represents the projection to the doublon-holon subspace of site $i$ spanned by $\ket{0}$ and $\ket{\uparrow\downarrow}$ and $\bar{\mathcal{P}}=1-\mathcal{P}$. This effective model is a generalization of the $t$--$J$ model, which is derived in the Appendix.
The two exchange interactions share the same coupling constant $J_{\rm ex}= 2t_0^2/U$, and thus are closely related. The $\eta$--exchange interaction originates from a virtual process exchanging a neighboring doublon-hole pair, see Appendix for more details. This model therefore explains both the antiferromagnetic phase at $d\sim0$ and the $\eta$--pairing at $d\sim0.5$. The above-mentioned universal photodoped state is, indeed, rigorously defined by this model. A photodoped Mott insulator is then characterized by a mixture of doublons/holons carrying $\eta$--pseudopin and localized electrons carrying spin. We further note that, in a chemically doped Mott insulator, only one type of the charge excitations (doublon or holon) exist and the $\eta$--exchange term $-J_{\rm ex}\eta^+_i\eta^-_j+\rm{h.c.}$ vanishes.

Furthermore, a particle-hole transformation $d_{i\uparrow}\to \tilde{d}_{i\uparrow},d_{i\downarrow}\to (-1)^i \tilde{d}^\dag_{i\downarrow}$ maps charge to spin ($\boldsymbol{\eta}_i\to\boldsymbol{\tilde{S}}_i$) and $U\to-U$. The sign reversal of $U$ is equivalent to the effect of negative temperature (a highly excited state with completely inverted charge distribution in the energy spectrum, see Ref.~\citenum{werner2019prb} for more details), leading to a negative temperature phase diagram which may be realized through strong external driving \cite{tsuji2011}, see the lower half of Fig.~\ref{pd}. Hence, the $\eta$--pairing phase is dual to a ferromagnetic state, with singly occupied sites mapped to charge excitations, which also explains its larger stability in terms of the stability of a ferromagnetic phase against defects. Indeed, the hopping of charge excitations in the FM phase does not create strings of defects in the ordered background, in contrast to the AFM phase, where this effect contributes significantly to the destruction of staggered spin ordering \cite{golez2014,grusdt2018}.

Finally, we also checked that the $\eta$--pairing phase survives in the presence of small (particle-hole) symmetry-breaking terms in the Hamiltonian, as one expects for a symmetry-breaking phase. 
In particular, all conclusions survive under a next-to-nearest-neighbor hopping $t_1=0.1t_0$. Different $U$ and $g$ are also studied, which give no qualitatively different results.

\begin{figure}
\includegraphics[scale=0.9]{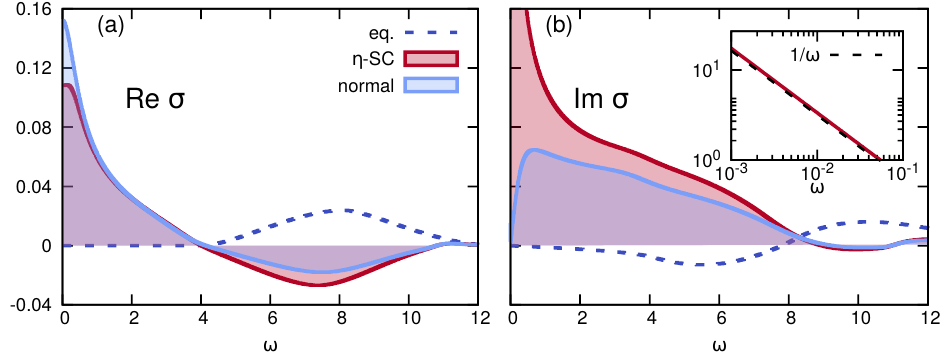}
\caption{(a) Real and (b) imaginary part of the optical conductivity. The dark blue dashed curves correspond to an equilibrium paramagnetic state with $\beta=100$. The light blue curves and the red curves characterize the two states labeled by red ($\eta$ state, $\beta_b$=100) and blue arrows (normal state, $\beta_b$=17.2) in Fig.~\ref{pd}, respectively. A delta function peak at $\omega=0$, present in the SC phase, is not shown. Both normal and $\eta$ states have $\mu_b=5.4$. For all three curves $\Gamma=0.05$ and $U=8.0$ are assumed. The inset shows a 1/$\omega$ scaling of the imaginary part of the optical conductivity.}
\label{cond}
\end{figure}

Since the bath coupling in the above discussion is weak and does not selectively favor the $\eta$--pairing phase, the latter should be an intrinsic property of the photodoped state, and thus be accessible in real time with any protocol which realizes strong photodoping at low $T_{\rm eff}$ and breaks the conservation of $\langle\boldsymbol{\eta}\rangle$. This is fundamentally different from the previous works requiring specific properties of the driving and the $SO(4)$ symmetry protection \cite{diehl2008,bernier2013,kitamura2016,kaneko2019,tindall2019}. We demonstrate the universality of our steady-state theory by considering two real-time protocols. Firstly, we consider the resonant excitation of doublon-hole pairs induced by an electric pulse in a Hubbard model coupled to bosonic baths, which leads to significantly enhanced $\chi_\eta=15$ up to the maximum simulation time. Better results can be obtained by coupling the Hubbard bands to external narrow bands, e.g., core levels, which cool down the electrons by absorbing large amounts of entropy (evaporative-cooling effect) \cite{werner2019}. In this case we observed a symmetry breaking $\eta$--paired phase which remains beyond $t\sim 100$ \cite{werner2019prb}, see Appendix~\ref{rt_prot} for more details.

\begin{figure}
\includegraphics[scale=0.7]{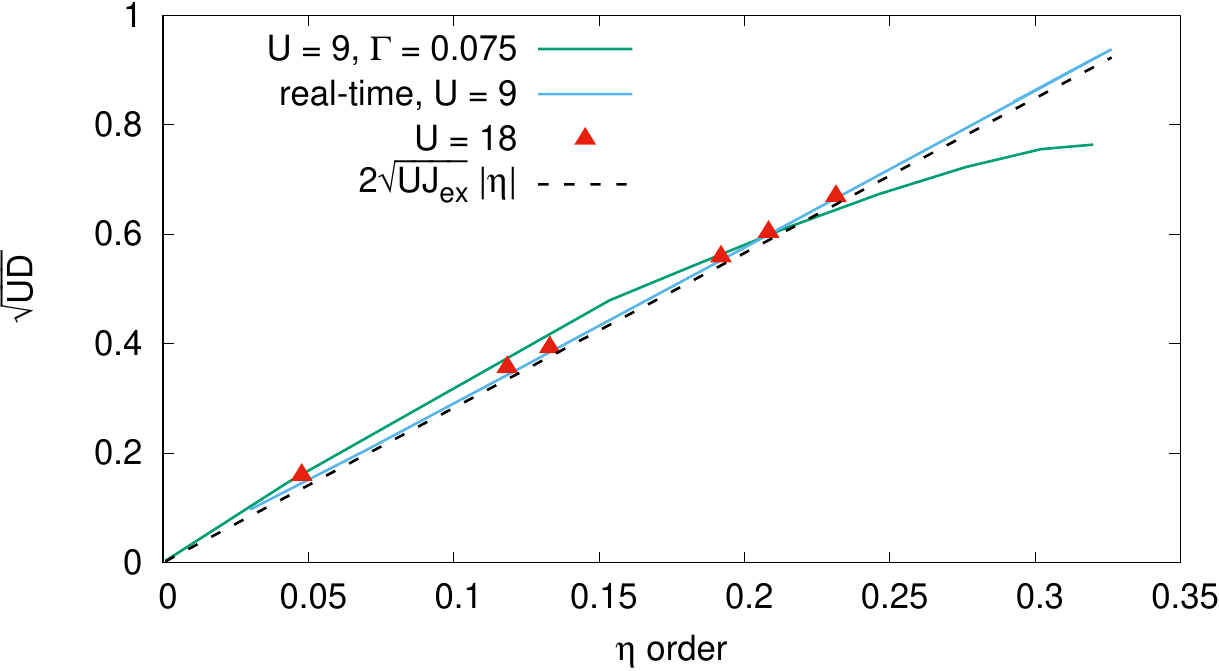}
\caption{$\sqrt{UD}$ as a function of $\eta$--pairing. The rescaling is intended to demonstrate the $D\propto\eta^2/U$ scaling. The real-time results for $U=9$ and $18$ are obtained using the entropy-cooling protocol (ii), where external narrow bands are coupled up to about $t\approx100$ and then detached, leaving behind an $\eta$--pairing phase. A superconducting current $j$ is then created by a short electric pulse satisfying $A=-\int dt E(t)\approx-0.020$ to measure $D$. The dashed line is predicted by the mean-field result for the phase stiffness.}
\label{drude}
\end{figure}

\subsection{Optical signature of the $\eta$--paired phase}
In this section, we study the superconducting optical response of the hidden phase. In DMFT, the optical conductivity can be evaluated from the current--current ($j$--$j$) correlation function $\chi_{jj}(t,t')=\delta j(t)/\delta A(t')$ with $\sigma(\omega)=-i\chi_{jj}(\omega)/(\omega+i0^+)$ in the steady-state \cite{eckstein2008}, see Appendix for detailed discussions. As shown in Fig.~\ref{cond}, the equilibrium system features a Mott gap of size $\sim U$ as expected. Strong photodoping with $d\sim0.4$ gives rise to a broad Drude peak in $\operatorname{Re}\sigma(\omega)$ ($\omega>0$), implying normal metallic behavior. More interestingly, a negative conductivity is observed at $\omega\sim U$ for $\operatorname{Re}\sigma$. This can be attributed to the recombination of doublon-hole pairs under periodic driving. The $\eta$--pairing phase is, on the other hand, characterized by a clear $1/\omega$ behavior in the imaginary part $\operatorname{Im}\sigma$ at small $\omega$, in contrast to the normal phase where $\operatorname{Im}\sigma(0)=0$. In this case, a delta function peak $\operatorname{Re}\sigma(\omega)\sim \pi D\delta(\omega)$ is imposed by analyticity (not shown in the plot), where we define the SC Drude weight $D=-\operatorname{Re}\chi_{jj}(0)$. This delta function peak leads to the zero resistivity effect. In addition, the zero-frequency \mbox{$j$--$j$} correlation $\chi_{jj}(0)\ne0$ results in the London equation $j=-D A$, where $D$ is identified with the phase stiffness $\langle \delta^2 H/\delta A^2\rangle$ \cite{yang1990,sewell1990}. 

We observe that the results obtained with different protocols (both real-time and steady-steate) for different doublon densities, different $\beta_{\rm eff}$ and different $U$ collapse onto a single line when $\sqrt{UD}$ is plotted against the order parameter $|\eta|$. (The deviation of the steady-state data at large $\eta$ may be attributed to non-thermal effects induced by the bath coupling.) This indicates a scaling behavior $D\sim |\eta|^2/U$ and can be explained by the two-liquid effective model. The phase stiffness can be evaluated to be $D=4J_{\rm ex}\langle \bm \eta_i\cdot\bm\eta_j\rangle\simeq4J_{\rm ex}|\eta|^2$ for neighboring sites $i,j$, see Appendix for a detailed derivation. The $\eta$--SC hidden phases are, therefore, universally %underscored 
supported 
by the instrinsic doublon-holon pairing mechanism. More interestingly, the phase stiffness corresponds to short-range correlations $\langle \bm \eta_i\cdot\bm\eta_j\rangle$, and may be observable under much smaller photodoping (recall that $\chi_\eta$ is strongly enhanced already for smaller $d$). 

\section{Conclusion}  
In this Article, we have studied the non-equilibrium phase diagram of the repulsive Hubbard model and demonstrated that the $\eta$--pairing superconducting phase can be stabilized by photodoping, i.e., the injection of cold photocarriers into the system. The hidden phase exists in a wide range of parameters ($d$, $\beta_{\rm eff}$) and requires no symmetry protection. 
Photodoping leads to normal metallic behavior and a negative conductivity at large frequencies, while the $\eta$--pairing phase is further characterized by a zero dc-resistivity and the Meissner effect. Moreover, the phase stiffness $D$ can be enhanced with increased short-range $\eta$ correlations, which may be ubiquitously observed in excited Mott insulators. Strong photodoping with $d>\frac{1}{4}$ in a Mott insulator can potentially be realized with recent proposals \cite{werner2019, peronaci2019}. A large density of doublon-hole pairs can also be prepared in fermionic cold-atom systems \cite{rosch2008,mazurenko2017,chiu2019}. The $\eta$--paired phase is observed here in an infinite-dimensional system, while, in lower dimensions, the strong enhancement of the pairing susceptibility can still have observable effects. The staggered $\eta$--pairing superconducting order can potentially be detected using a recently proposed noise correlation measurement in ARPES experiments \cite{stahl2019}.  

Independent of the experimental implementation, our main finding of an $\eta$--pairing doublon-hole condensate over a broad range of doping levels is of general importance for various Mott insulators. Our approach to prepare cold photodoped states may be used to explore related unconventional SC orders in more complex Mott insulators \cite{kugel1982,pesin2010}, and in charge transfer insulators \cite{golez2019}.  

\begin{acknowledgments}
We acknowledge discussions with T. Kaneko, O. Parcollet, and A. Millis. M.E. and J. Li were supported by the ERC starting grant No. 716648. PW acknowledges support from ERC Consolidator Grant No. 724103. The Flatiron institute is a division of the Simons foundation.
\end{acknowledgments}
\appendix
%\section{Appendix}
\section{Dynamical Mean-Field Theory}
\label{Sdmft}
\begin{figure}
\includegraphics[scale=1.5]{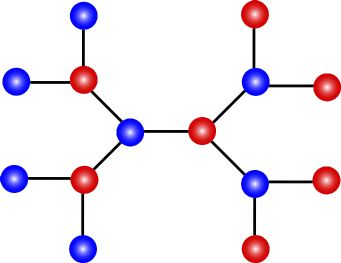}
\caption{An example of Bethe lattice with coordination number $3$. The two sublattices are distinguished by red and blue colors.}
\end{figure}
This section provides details on the non-equilibrium formulation of Dynamical Mean-Field Theory used in the main text. The simulation is done on a Bethe lattice with an inifinite coordination number. A Bethe lattice is a tree-like structure without loops. It can be naturally decomposed to two sublattices, thus is available for treating staggered orders in bipartite lattices, such as the antiferromagnetic order in a square lattice. To deal with the superconducting order, we define the Nambu spinor $\psi^T=(\psi_\uparrow,\psi_\downarrow)=(d_\uparrow,d^\dag_\downarrow)$ and the Hamiltonian is rewritten as,
\begin{align}
H=-t_0\sum_{\langle ij\rangle\sigma} \sigma{\rm e}^{i\sigma \mathcal{A}}\psi^\dag_{i\sigma} \psi_{j\sigma}-U\sum_{i}\psi^\dag_{i\uparrow}\psi_{i\uparrow} \psi^\dag_{i\downarrow}\psi_{i\downarrow},
\end{align}
where the coupling to a general vector potential $\mathcal{A}(t)$ is assumed. The lattice problem is mapped to a single-impurity Anderson model, which is defined by the following action,
\begin{align}
S_{A/B}[\psi,\bar{\psi}]=S^{\rm loc}_{A/B}[\psi,\bar{\psi}]-\int dt dt' \bar{\psi}(t)\Delta_{A/B}(t,t') \psi(t'),
\end{align}
where $S^{\rm loc}$ collects the local terms and $\Delta_{A,B}$ is the matrix-valued hybridization function determined by the self-consistency condition for a Bethe lattice of infinite coordination number \cite{georges1996},
\begin{align}
\Delta_{A/B}(t,t')&=\Delta_{A/B}^R(t,t')+\Delta_{A/B}^L(t,t')+D(t,t'),\text{ with}\nonumber\\
\Delta_{A/B}^R(t,t')&=\frac{t_0^2}{2}\sigma_z {\rm e}^{i\sigma_z\mathcal{A}(t)}G_{B/A}(t,t'){\rm e}^{-i\sigma_z\mathcal{A}(t')}\sigma_z,\nonumber\\
\Delta_{A/B}^L(t,t')&=\frac{t_0^2}{2} \sigma_z{\rm e}^{-i\sigma_z\mathcal{A}(t)}G_{B/A}(t,t'){\rm e}^{i\sigma_z\mathcal{A}(t')}\sigma_z,
\end{align}
where $D=D^++D^-$ comes from the bath coupling as discussed in the main text. $G_{A/B}$ are matrix-valued impurity Green's function in Nambu basis. We assume half of the bonds connected the local site are parallel to the external field $\mathcal{A}$ while the other half are antiparallel to it, leading to Peierls phases of sign $\pm1$, respectively. $\mathcal{A}(t)$ is set to zero in the bath-doping and entropy transfer protocols (protocol (ii) in main text). The two sublattices $A,B$ are related by $G_B=\sigma_z G_A\sigma_z$ in the presence of $\eta$--pairing and with $\sigma_z$ replaced by the identity matrix in the $s$--wave pairing case.

One can also consider the next-nearest-neighbor (NNN) hopping $-t_1\sum_{\langle\langle ij\rangle\rangle\sigma}\sigma{\rm e}^{i\sigma \mathcal{A}}\psi^\dag_{i\sigma} \psi_{j\sigma}$. This is included by adding $\frac{t_1^2}{2}\sigma_z {\rm e}^{i2\sigma_z\mathcal{A}(t)}G_{A/B}(t,t'){\rm e}^{-i2\sigma_z\mathcal{A}(t')}\sigma_z$ to hybridization function $\Delta^R_A$ and analogously for $\Delta^L$ and sublattice $B$.

\section{The two-liquids effective model of photodoped states}
The low energy physics of a fermionic Hubbard model is described by a $t$--$J$ model under hole or electron doping \cite{macdonald1988,bukov2016}. In a photodoped state, the situation is different due to the simultaneous presence of nonthermal doublons and holes. To derive the effective theory of the photodoped state, we assume the double occupancy is conserved, which is justified by an exponentially large lifetime of doublon-hole pairs as discussed in the main text. We then perform a Schrieffer-Wolff transformation following Ref. \citenum{bukov2016}. Specifically, we transform the Hamiltonian~(1) into a rotating frame by unitary transformation $H^{\rm rot}(t)={\rm e}^{S(t)}(H-i\partial_t){\rm e}^{-S(t)}$, with $S(t)=Ut\sum_i(n_{i\uparrow}-1/2)(n_{i\downarrow}-1/2)$. The resulting Hamiltonian reads as follows,
\begin{align}
H^{\rm rot}(t)&=-t_0\sum_{\langle ij\rangle\sigma}[\overline{\mathcal{P}}_i d^\dag_{i\sigma}d_{j\sigma}\overline{\mathcal{P}}_j+\mathcal{P}_i d^\dag_{i\sigma}d_{j\sigma}\mathcal{P}_j]-\nonumber\\
&-t_0\sum_{\langle ij\rangle\sigma}[{\rm e}^{iUt}\mathcal{P}_i d^\dag_{i\sigma}d_{j\sigma}\overline{\mathcal{P}}_j+{\rm h.c.}],
\label{hrot}          
\end{align}
where $\mathcal{P}_i=n_{i\uparrow}n_{i\downarrow}+\bar{n}_{i\uparrow}\bar{n}_{i\downarrow}=1-(n_{i\uparrow}+n_{i\downarrow})+2n_{i\uparrow}n_{i\downarrow}$ projecting site $i$ to the doublon-hole subspace with $\bar{n}=1-n$ and $\overline{\mathcal{P}}=1-\mathcal{P}$. The first line of \eqref{hrot} switches a doublon/hole state with its neighboring singlon state, while the second line gives rise to the creation and recombination of doulon-hole pairs. In this formalism, both creation and recombination processes are treated on equal footing and can be integrated out through a high-frequency expansion \cite{bukov2015}. Ignoring three-site terms, the following results can be checked with straightforward calculations \cite{peronaci2019},
\begin{align}
H^{\rm eff}&=-t_0\sum_{\langle ij\rangle\sigma}[\overline{\mathcal{P}}_id^\dag_{i\sigma}d_{j\sigma}\overline{\mathcal{P}}_j+\mathcal{P}_i d^\dag_{i\sigma}d_{j\sigma}\mathcal{P}_j]+\nonumber\\
&+\frac{t_0^2}{U}\sum_{\langle ij\rangle\sigma}[d^\dag_{i\sigma}d_{i\bar{\sigma}}d^\dag_{j\bar{\sigma}} d_{j\sigma}+d^\dag_{i\sigma}d^\dag_{i\bar{\sigma}}d_{j\bar{\sigma}}d_{j\sigma}]\nonumber\\
&+\frac{t_0^2}{U}\sum_{\langle ij\rangle\sigma}(n_{i\sigma}-n_{j\sigma})n_{i\bar{\sigma}}\bar{n}_{j\bar{\sigma}}.
\label{heff}
\end{align}
The first term in the second line is simply $S^+_iS^-_j+S^-_iS^+_j$. Using the identity $\sum_{\langle ij\rangle}a_{ij}=\sum_{\langle ij\rangle}a_{ji}$, the second term of the line can be rewritten by
\begin{align}
&\frac{1}{2}\sum_{\langle ij\rangle\sigma}[d^\dag_{i\sigma}d^\dag_{i\bar{\sigma}}d_{j\bar{\sigma}}d_{j\sigma}+d^\dag_{j\sigma}d^\dag_{j\bar{\sigma}}d_{i\bar{\sigma}}d_{i\sigma}]\nonumber\\
&=-\sum_{\langle ij\rangle}[\eta^+_i\eta^-_j+\eta^-_i\eta^+_j].
\end{align}
The third line of \eqref{heff} can be simplified by projecting into doublon-hole and singlon subspaces, defining $F_{ij\sigma}=(n_{i\sigma}-n_{j\sigma})n_{i\bar{\sigma}}\bar{n}_{j\bar{\sigma}}$,
\begin{align}
F_{ij\sigma}&=\mathcal{P}_i F_{ij\sigma}\mathcal{P}_j+\overline{\mathcal{P}}_i F_{ij\sigma}\mathcal{P}_j\nonumber\\
&+\mathcal{P}_i F_{ij\sigma}\overline{\mathcal{P}}_j+\overline{\mathcal{P}}_i F_{ij\sigma}\overline{\mathcal{P}}_j.
\end{align}
It can be checked that the second and third terms identically vanish. The first and fourth terms read
\begin{align}
\sum_{\langle ij\rangle\sigma}\mathcal{P}_i F_{ij\sigma}\mathcal{P}_j&=\sum_{\langle ij\rangle}\mathcal{P}_i (n_{i\uparrow}\bar{n}_{j\downarrow}+n_{i\downarrow}\bar{n}_{j\uparrow})\mathcal{P}_j,\nonumber\\
&=-\frac{1}{2}\sum_{\langle ij\rangle}\mathcal{P}_i[(n_{i\uparrow}+n_{i\downarrow}-1)(n_{j\uparrow}+n_{j\downarrow}-1)\nonumber\\
&\quad-1]\mathcal{P}_j\nonumber\\
&=\sum_{\langle ij\rangle}-2\eta_i^z\eta_j^z+\frac{1}{2}\mathcal{P}_i\mathcal{P}_j,\\
\sum_{\langle ij\rangle\sigma}\overline{\mathcal{P}}_i F_{ij\sigma}\overline{\mathcal{P}}_j&=-\sum_{\langle ij\rangle}\overline{\mathcal{P}}_i(n_{j\uparrow}n_{i\downarrow}+n_{j\downarrow}n_{i\uparrow})\overline{\mathcal{P}}_j\nonumber\\
&=\frac{1}{2}\sum_{\langle ij\rangle}\overline{\mathcal{P}}_i[(n_{i\uparrow}-n_{i\downarrow})(n_{j\uparrow}-n_{j\downarrow})-1]\overline{\mathcal{P}}_j\nonumber\\
&=\sum_{\langle ij\rangle}2S^z_iS^z_j-\frac{1}{2}\overline{\mathcal{P}}_i\overline{\mathcal{P}}_j,
\end{align}
where $\overline{\mathcal{P}}_i n_i=1$ and $\mathcal{P}_i n_{i\uparrow}=\mathcal{P}_i n_{i\downarrow}$ are used. Collecting these terms, one then reaches the following form
\begin{widetext}
\begin{align}
H^{\rm eff}=-\sum_{\langle ij\rangle}J_{\rm ex}\boldsymbol{\eta}_i\cdot \boldsymbol{\eta}_j+\sum_{\langle ij\rangle}J_{\rm ex}\boldsymbol{S}_i\cdot \boldsymbol{S}_j-t_0\sum_{\langle ij\rangle\sigma}[\overline{\mathcal{P}}_id^\dag_{i\sigma}d_{j\sigma}\overline{\mathcal{P}}_j+\mathcal{P}_i d^\dag_{i\sigma}d_{j\sigma}\mathcal{P}_j]+\frac{J_{\rm ex}}{4}\sum_{\langle ij\rangle}(\mathcal{P}_i\mathcal{P}_j-\overline{\mathcal{P}}_i\overline{\mathcal{P}}_j).
\label{heff_f}
\end{align}
\end{widetext}
The first and second terms represent the $\eta$--$\eta$ exchange interaction between doublon-hole pairs and the regular superexchange interaction between spins, respectively. The third term represents hopping of electrons and couples doublon-hole and singlon liquids. It is unchanged from \eqref{heff}. The factor $\hat{M}=\sum_{\langle ij\rangle}(\mathcal{P}_i\mathcal{P}_j-\overline{\mathcal{P}}_i\overline{\mathcal{P}}_j)$ can be simplified to
\begin{align}
\hat{M}&=\sum_{\langle ij\rangle}(-1+\mathcal{P}_i+\mathcal{P}_j)\nonumber\\
&=\sum_{\langle ij\rangle}(2(n_{i\uparrow}n_{i\downarrow}+n_{j\uparrow}n_{j\downarrow})-n_i-n_j+1)\nonumber\\
&=4D\sum_i \left(n_{i\uparrow}-\frac{1}{2}\right)\left(n_{i\downarrow}-\frac{1}{2}\right)+{\rm const.},
\end{align}
where $D$ is the coordination number of the lattice. This is a constant if double occupancy and total particle number are fixed and can, therefore, be neglected in our discussion. Notice that the Hamitonian~\eqref{heff_f} does not depend on the half-filling condition and becomes equivalent to the $t$--$J$ model in the absence of either doublons or holes. In fact, the absence of doublons or holes makes $\eta^+=\eta^-=0$ and $J_{\rm ex}\boldsymbol{\eta}_i\cdot \boldsymbol{\eta}_j\to J_{\rm ex} n_i n_j/4$. A finite chemical potential $\mu$ results in the Zeeman term $\mu\eta_z$ breaking the charge-sector $SU(2)$ symmetry. Moreover, next-nearest-neighbor (NNN) hopping results in frustration in the ordering. These terms, when being small, should suppress but not necessarily wipe out the $\eta$--pairing phase completely.

\subsection{Phase stiffness of the $\eta$--pairing condensate}
A vector potential coupled to the model \eqref{heff_f} can generally result in modification of the parameter $J_{\rm ex}$ \cite{eckstein2017}. However, for slowly varying fields, especially for a constant vector potential $\mathbf{A}$, a simpler treatment is available as follows. Consider a half-filled repulsive Hubbard model which couples to a constant and spatially uniform vector potential $\mathbf{A}$. 

\begin{align} 
H=-t\sum_{\langle ij\rangle}{\rm e}^{i \mathbf{A}\cdot(\mathbf{R}_i-\mathbf{R}_j)}d^\dag_{i\sigma} d_{j\sigma}+U\left(n_\uparrow-\frac{1}{2}\right)\left( n_\downarrow-\frac{1}{2}\right).
\end{align}

This coupling can clearly be gauged away (absorbed by a gauge transformation), and thus is physically irrelevant in the non-superconducting phase,
\begin{align}
d_{i\sigma}={\rm e}^{i\mathbf{A}\cdot\mathbf{R}_i}\tilde{d}_{i\sigma}.
\end{align}
However, the $\eta$--pseudospin operators transform under the transformation
\begin{align}
\eta^+_i=\eta^x_i+i\eta^y_i=\theta_i d^\dag_{i\uparrow}d^\dag_{i\downarrow}={\rm e}^{-i2\mathbf{A}\cdot\mathbf{R}_i}\tilde{\eta}^+_i,
\end{align}
which applies to $\eta^{-}=\eta^{+\dag}$ analogously. As mentioned above, one can generally consider a time-dependent vector potential $\mathbf{A}(t)$ and make a time-dependent Schrieffer-Wolff transformation. Here we restrict ourselves to a constant $\mathbf{A}(t)=\mathbf{A}$, in which case the effective model would simply be a Heisenberg-like model. With $J_{\rm ex}=2t^2/U$,
\begin{align}
H_{\rm eff}&=-\frac{J_{\rm ex}}{2}\sum_{\langle ij\rangle}(\tilde{\eta}^+_i\tilde{\eta}^-_j+{\rm h.c.})+H_{\rm hop}\nonumber\\
&=-\frac{J_{\rm ex}}{2}\sum_{\langle ij\rangle}({\rm e}^{i2\mathbf{A}\cdot(\mathbf{R}_i-\mathbf{R}_j)}\eta^+_i \eta^-_j+{\rm h.c.})+H_{\rm hop},
\end{align}
which demonstrates the coupling between $\eta$--pseudospins and the gauge field. Note that $H_{\rm hop}$ is the doublon/hole hopping term giving rise to the regular part of optical conductivity. With normal current neglected, the SC current along bond $\mathbf{e}=\mathbf{R}_j-\mathbf{R}_i$ can be expressed as 
\begin{align}
j_e&=-\langle\delta H_{\rm eff}/\delta A_e\rangle\nonumber\\
&=iJ_{\rm ex}\langle {\rm e}^{2iA_e}\eta^+_{i}\eta^-_{j}-{\rm h.c.}\rangle\nonumber\\
&=2J_{\rm ex}[-\langle \boldsymbol{\eta}_{i}\times\boldsymbol{\eta}_{j}\rangle_z\cos(2A_e)+\langle \boldsymbol{\eta}_{i}\cdot\boldsymbol{\eta}_{j}\rangle\sin(2A_e)]
\label{je}
\end{align}
where $A_e=\mathbf{A}\cdot\mathbf{e}$ and $\eta^z=0$ are assumed. 

For weak $A_e$, it is conventional to calculate the phase stiffness $D=-\langle \delta^2 H_{\rm eff}/\delta^2 A_e\rangle|_{A_e=0}$, which appears
in the London equation $j_e = -D A_e$ as usual, 
\begin{align}
D=4J_{\rm ex}\langle \bm \eta_i \cdot \bm \eta_j\rangle.
\end{align}
In the $\eta$--pairing phase, the relatively large $|\eta|$ order justifies a mean-field approximation, which yields $D=4J_{\rm ex}|\eta|^2$.

\section{Optical conductivity in Bethe lattice}
The optical conductivity can be computed within DMFT using single-particle quantities \cite{eckstein2008}. A lattice summation is usually required in the calculation. However, the procedure can be much simplified in the case of a Bethe lattice. In this section we derive the longitudinal optical conductivity in the Bethe lattice. Under a spatially uniform electric-field, the hybridization function is $\Delta_A(t,t')=\Delta_A^R(t,t')+\Delta_A^L(t,t')$, where we have assumed two sublattices $A,B$ as above. In the following we will omit the subscript $A,B$ unless it would be ambiguous. Then the current can be expressed as follows,
\begin{widetext}
\begin{align}
J(t)&=-\frac{1}{2}\operatorname{Re}\operatorname{Tr}\left(\sigma_z G\circ(\Delta_R-\Delta_L)\right)^<\nonumber\\
&=-\frac{1}{2}\operatorname{Re}\operatorname{Tr}\left\{\int_{-\infty}^t ds \sigma_z G^r(t,s)\left(\Delta^<_R(s,t)-\Delta^<_L(s,t)\right)+\int_{-\infty}^t ds \sigma_z G^<(t,s)\left(\Delta^a_R(s,t)-\Delta^a_L(s,t)\right)\right\}.
\end{align}
\end{widetext}
To obtain the susceptibility, we differentiate the functional $J[\mathcal{A}(t)]$
\begin{widetext}
\begin{align}
\chi(t,t')&=\delta J_A(t)/\delta \mathcal{A}(t')\rvert_{\mathcal{A}=0}\nonumber\\
&=-\frac{1}{2}\operatorname{Re}\operatorname{Tr}\left\{\int_{-\infty}^t ds \sigma_z\frac{\delta G^r(t,s)}{\delta \mathcal{A}(t')}\left(\Delta^<_R(s,t)-\Delta^<_L(s,t)\right)+\int_{-\infty}^t ds \sigma_z\frac{\delta G^<(t,s)}{\delta \mathcal{A}(t')}\left(\Delta^a_R(s,t)-\Delta^a_L(s,t)\right)\right\}\bigg\rvert_{\mathcal{A}=0}-\nonumber\\
&-\frac{1}{2}\operatorname{Re}\operatorname{Tr}\left\{\int_{-\infty}^t ds \sigma_zG^r(t,s)\left(\frac{\delta \Delta^<_R(s,t)}{\delta \mathcal{A}(t')}-\frac{\delta \Delta^<_L(s,t)}{\delta \mathcal{A}(t')}\right)+\int_{-\infty}^t ds \sigma_zG^<(t,s)\left(\frac{\delta \Delta^a_R(s,t)}{\delta \mathcal{A}(t')}-\frac{\delta \Delta^a_L(s,t)}{\delta \mathcal{A}(t')}\right)\right\}\bigg\rvert_{\mathcal{A}=0}.
\label{chi1}
\end{align}
\end{widetext}
We note that $\Delta_R=\Delta_L$ for $\mathcal{A}(t)=0$. This causes the first line of \eqref{chi1} to vanish identically. On the other hand, the functional derivative of $\Delta$ leads to
\begin{widetext}
\begin{align}
\frac{\delta \Delta^{<,a}_R(s,t)}{\delta \mathcal{A}(t')}&=\frac{t_0^2}{2}{\rm e}^{i\sigma_z\mathcal{A}(s)}\sigma_z\left\{\frac{\delta G^{<,a}(s,t)}{\delta \mathcal{A}(t')}+i\delta(s-t')\sigma_zG^{<,a}(s,t)-i\delta(t-t')G^{<,a}(s,t)\sigma_z\right\}\sigma_z{\rm e}^{-i\sigma_z\mathcal{A}(t)},\nonumber\\
\frac{\delta \Delta^{<,a}_L(s,t)}{\delta \mathcal{A}(t')}&=\frac{t_0^2}{2}{\rm e}^{-i\sigma_z\mathcal{A}(s)}\sigma_z\left\{\frac{\delta G^{<,a}(s,t)}{\delta \mathcal{A}(t')}-i\delta(s-t')\sigma_zG^{<,a}(s,t)+i\delta(t-t')G^{<,a}(s,t)\sigma_z\right\}\sigma_z{\rm e}^{i\sigma_z\mathcal{A}(t)},
\end{align}
\end{widetext}
where the $\delta G/\delta \mathcal{A}$ terms cancel on subtracting the two terms at $\mathcal{A}=0$. Thus we finally have
\begin{widetext}
\begin{align}
\chi_A(t,t') &= -\frac{t_0^2}{2}\operatorname{Re}\operatorname{Tr}\bigg\{i\int_{-\infty}^t ds \sigma_z G^r_A(t,s)\sigma_z\left(\sigma_zG^<_B(s,t)\delta(s-t')-G^<_B(s,t)\sigma_z\delta(t-t')\right)\sigma_z+\nonumber\\
&+i\int_{-\infty}^t ds \sigma_z G^<_A(t,s)\sigma_z\left(\sigma_z G^a_B(s,t)\delta(s-t')-G^a_B(s,t)\sigma_z\delta(t-t')\right)\sigma_z\bigg\}\nonumber\\
&=-\frac{t_0^2}{2}\operatorname{Im}\operatorname{Tr}\sigma_z\bigg\{G_A^r(t,t') G_B^<(t',t)\sigma_z+G^<_A(t,t') G^a_B(t',t)\sigma_z-\nonumber\\
&-\delta(t-t')\int_{-\infty}^\infty ds \left(G_A^r(t,s)\sigma_zG_B^<(s,t)+G_A^<(t,s)\sigma_zG_B^a(s,t)\right)\bigg\}.
\end{align}
\end{widetext}
%JL
The formula only contains local Green's functions and all $\delta G/\delta \mathcal{A}$ like terms are cancelled, so no lattice summation is required for computing $\chi$. To study the normal phase, we can go back to the original formalism (in contrast to Nambu formalism) by simply replacing $\sigma_z$ with the identity matrix $\mathbb{I}$. It is straightforward to verify that the Drude weight vanishes. 
In fact,
\begin{widetext}
\begin{align}
D(t)&=-\int_{-\infty}^{\infty}dt'\chi(t,t')\nonumber\\
&=\frac{t_0^2}{2}\operatorname{Im}\operatorname{Tr}\bigg\{\int_{-\infty}^\infty dt' \left(G_A^r(t,t')G_B^<(t',t)\mathbb{I}+G_A^<(t,t')G_B^a(t',t)\mathbb{I}\right)-\int_{-\infty}^\infty dt' \left(G_A^r(t,t')\mathbb{I}G_B^<(t',t)+G_A^<(t,t')\mathbb{I}G_B^a(t',t)\right)\bigg\}\nonumber\\
&=0
\end{align}
\end{widetext}

\subsection{f-sum rule}
$\sigma(t,t')=-c\int_{t'}^{\infty}d\bar{t}\chi(t,\bar{t})$ has a jump at $t=t'$ due to the delta function in $\chi$. As a result, the integration of $\sigma(t,\omega)$ satisfies $\int_0^\infty d\omega \sigma(t,\omega)=\frac{1}{2}\int_{-\infty}^\infty d\omega \sigma(t,\omega)=\frac{1}{4}2\pi\sigma(t,t^-)$ and 
\begin{widetext}
\begin{align}
\int_0^\infty d\omega \sigma(t,\omega)&=-\frac{\pi t_0^2}{4}\operatorname{Im}\operatorname{Tr}\bigg\{\sigma_z\int_{-\infty}^\infty ds \left(G_A^r(t,s)\sigma_zG_B^<(s,t)+G_A^<(t,s)\sigma_zG_B^a(s,t)\right)\bigg\}\nonumber\\
&=-\frac{\pi}{4}\operatorname{Im}\operatorname{Tr}\bigg\{\int_{-\infty}^\infty ds \left(G_A^r(t,s)(t_0^2\sigma_z G_B^<(s,t)\sigma_z)+G_A^<(t,s)(t_0^2\sigma_zG_B^a(s,t)\sigma_z)\right)\bigg\}\nonumber\\
&=-\frac{\pi}{4}E_{\rm kin}.
\end{align}
\end{widetext}
%JL
This is consistent with the sum rule in more realistic lattices \cite{aoki2014}, and thus justifies the use of Bethe lattice to study the optical conductivity.
\section{Real-time protocols}
\label{rt_prot}
In addition to the steady-state results obtained with fermion-reservoir coupling, we also considered two real-time protocols to demonstrate the universality of photodoped states and crosscheck the observable properties computed from different ways. An obvious attempt to reach a cold photodoped state is to create doublons and holes by resonant excitation between the Hubbard bands, and subsequently ``cool'' them through coupling to a bath of bosonic degrees of freedom (phonons, spins) \cite{eckstein2013}. However, direct real-time simulations of this process \cite{eckstein2013, peronaci2019} close to the Mott state have so far reported only relatively high $T_{\rm eff}$. DMFT studies also suggest that the preparation of low temperature phases, such as a Fermi liquid state, by cooling from a hot state is critically slowed down by fundamental constraints apparently independent of the bath setting \cite{sayyad2019}.

To prepare the $\eta$-state it therefore seems advantageous to keep $T_{\rm eff}$ low throughout the process in which doublons and holes are being created.

\subsection{The evaporative cooling protocol}
In this protocol, we consider the scenario where an electric pulse couples the Hubbard bands with some narraow energy bands in the system, resulting in dipolar excitation between the system and the external bands. Specifically, we model this process as following. The single-band Hubbard model (in the Mott insulating phase) is suddenly coupled to two narrow-band fermion reservoirs. Doublons are created in the upper Hubbard band by coupling to a narrow full band, while the lower Hubbard band is emptied by ejecting singly occupied states into the narrow empty band. Doublons and holes are cooled by entropy transfer to the narrow bands, whose width controls the effective temperature \cite{werner2019}. These reservoirs are detached from the system after a short period (usually about $100$ hopping times). 

The driving-induced fermionic coupling results in a contribution to the hybridization function \cite{werner2019prb}
\begin{align}
\Delta_{\rm coupl}=\sum_\alpha v(t)G^0_{{\rm bath},\alpha}(t,t')v(t'),
\end{align}
where $\alpha=1,2$ indicates the two reservoirs and $G^0$ is the reservoir Green's function in equilbrium. The coupling $v(t)=f(t)\sin(\Omega(t)t)$ has generically a slowly-varying frequency $\Omega(t)$ and is modulated by the envelope function $f(t)$. In the adiabatic limit of slowly varying $\Omega(t)$, the density of states of the external fermionic bands is effectively shifted by $\pm\Omega(t)$ at time $t$, which leads to particle flux when the DoS of external baths overlaps with the upper or lower Hubbard bands. We consider a chirped pulse as follows
\begin{align}
\omega(t)=\omega_{0}+\Delta \omega\sin\left(\frac{\pi}{2}\frac{t}{t_{\rm ramp}}\right)\quad t<t_{\rm max},
\end{align}
where $\omega_0$ is selected so that doublons are initially ($t=0$) being created at the bottom of the upper Hubbard band, and $\Delta\omega$ is chosen to be close to the bandwidth making the upper band strongly populated at the end of $t_{\rm max}\approx 100$, after which the coupling is turned off. 

For obtaining the data points shown in Fig.~4 in the main text, we have chosen $\omega_0=8,\delta\omega=4.5,t_{\rm ramp}=170$ for the case $U=9$. The narrow bands are located at energy $\omega=\pm6$ and have box-shaped DoS with bandwidth $0.05$. More details can be found in Ref.~\cite{werner2019prb}.

\subsection{The direct excitation protocol}
In this section, we will consider a direct dipolar exctiation between lower and upper Hubbard bands through an electric pulse. For large enough gap, this protocol leads to a nonequilibrium (quasi)-steady state. This analysis is an extension of the work done in Ref.~\citenum{peronaci2019} to the symmetry-broken phase. The authors of Ref.~\citenum{peronaci2019} considered a periodic driving and a coupling to a continuum of phonons. We have repeated their analysis within the symmetry-broken formalism and applied a pairing field. Despite an exhaustive scan over the parameter space, we could not find a symmetry broken state in the long-time limit. The main obstacle is a high effective temperature, which was typically $\beta_{\rm eff}\approx 1$,  in agreement with the analysis in Ref.~\citenum{peronaci2019}. Here, we propose an alternative protocol using a chirped electric field pulse. In this protocol, the electric field is given by $E(t)=E_0 \sin(\omega(t) t),$ where we use a slow chirping of the electric field $\omega(t)=\omega_0+\alpha t.$  This protocol leads to a build-up of  distributions with lower effective temperatures as does the periodic driving, but the minimum rate of the chirping is limited by the maximum propagation time accessible in the numerics. 

\begin{figure}
\includegraphics[scale=1.0]{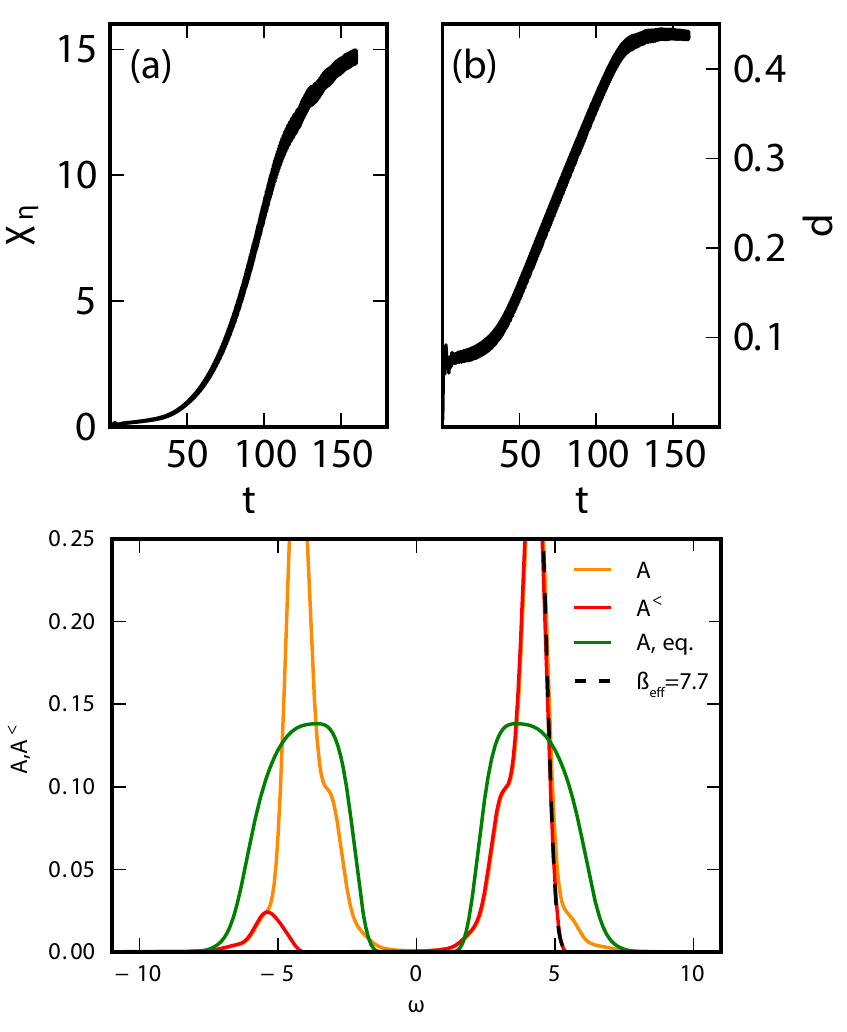}
\caption{Time evolution of the susceptibility $\chi_{\eta}$~(a) and double occupancy~(b) for a system excited by a chirped electric field and coupled to a phononic bath. (c) Long-time averaged spectral function $A(\omega,t=130)$~(orange) and the lesser component $A^<(\omega,t=130)$~(red). The dashed line indicates the Fermi distribution function for the inverse temperature $\beta_{}$ corresponding to the effective temperature $\beta_{\rm eff}=7.7$. The chirping rate was $\alpha=0.015$, the frequency of the high-energy $\omega_0=1.0$ and the low-energy $\omega_0=0.2$ and the electron-phonon coupling $\lambda=0.3$.}
\label{pair_time}
\end{figure}

The most efficient cooling bath that we found is a combination of a high-energy $\omega_{H}=1.0$ and low-energy $\omega_H=0.2$ phonon. To induce a symmetry breaking, we have applied a weak pairing field $h_x=0.01$, which we gradually turn off, and follow the time evolution of the pairing susceptibility $\chi_{\eta}$, see Fig.~\ref{pair_time}(a). The pairing susceptibility is strongly increased. Similarly,  the double occupancy $d$ is strongly enhanced, see Fig.~\ref{pair_time}(b), and in the long-time limit reaches the value $d=0.44.$ Due to the applied electric field, we have averaged the spectral function $A(\omega, t)=-\frac{1}{\pi} \mathrm{Im}\int_t^{t+t_{\text{cut}}} d{t'} e^{i\omega(t'-t)} G^r(t', t)$ over two periods of the oscillation $\bar A(\omega, t)=\frac{1}{T}\int_{t-T/2}^{t+T/2} A(\omega, t) dt,$ where $T=4\pi \omega (t)$, see Fig.~\ref{pair_time}(c). By fitting to a Fermi distribution function in the upper Hubbard band, we have determined the effective inverse temperature to be $\beta_{\rm eff}=7.7.$ A comparison with the phase diagram in Fig.~2 would suggest that a state with double occupancy $d=0.44$ and $\beta_{\rm eff}=7.7$ should be within the symmetry broken state.  This is further confirmed by the fact that the order parameter persists even when the pairing field $h_x$ is turned off. However, a strict criterion for the spontaneous symmetry breaking can be that the final state becomes independent of the size of the initial pairing field $h_x.$ A test with $h_x=0.001$ shows that the state in the long-time limit still depends on the initial pairing field. Despite the strongly enhanced susceptibility $\chi_{\eta}\sim15$ in the long-time limit, the dependence on the initial pairing field implies that the reached state has not yet entered the $\eta$-pairing state. However, the $\chi_\eta$ keeps increasing for the longest propagation times available and may eventually lead to a symmetry-broken phase for significantly longer simulation times.

\bibliography{eta.bib}

\end{document}